\documentclass[conference]{IEEEtran}
\IEEEoverridecommandlockouts
\usepackage{cite}
\usepackage{amsmath,amssymb,amsfonts}
\usepackage{algorithmic}
\usepackage{graphicx}
\usepackage{textcomp}
\usepackage{xcolor}

\usepackage{float}
\usepackage{textcomp}
\usepackage[ruled,linesnumbered]{algorithm2e}
\usepackage{booktabs} % For better looking tables
\usepackage{subfigure}
\usepackage{subcaption} % For subfigures
\usepackage{url}
\usepackage{stfloats}
\usepackage{balance}
\usepackage{multirow}
\def\BibTeX{{\rm B\kern-.05em{\sc i\kern-.025em b}\kern-.08em
    T\kern-.1667em\lower.7ex\hbox{E}\kern-.125emX}}
\begin{document}

\title{EAQGA: A Quantum-Enhanced Genetic Algorithm with Novel Entanglement-Aware Crossovers}

\author{
    Mohammad Kashfi Haghighi, Matthieu Fortin-Deschênes, Christophe Pere, Mickaël Camus\\
    \small $\textit{PINQ}^\textit{2}$\\
    \small \texttt{\{mohammad.kashfi, matthieu.fortin-deschenes, christophe.pere, mickael.camus\}@pinq2.com}
}

\maketitle

\renewcommand{\thefootnote}{}
\footnotetext{
    \textit{This work has been submitted to the IEEE for possible publication. 
    Copyright may be transferred without notice, after which this version may no longer be available.}
}

\begin{abstract}
Genetic algorithms are highly effective optimization techniques for many computationally challenging problems, including combinatorial optimization tasks like portfolio optimization. Quantum computing has also shown potential in addressing these complex challenges. Combining these approaches, quantum genetic algorithms leverage the principles of superposition and entanglement to enhance the performance of classical genetic algorithms. In this work, we propose a novel quantum genetic algorithm introducing an innovative crossover strategy to generate quantum circuits from a binary solution. We incorporate a heuristic method to encode entanglement patterns from parent solutions into circuits for the next generation. Our algorithm advances quantum genetic algorithms by utilizing a limited number of entanglements, enabling efficient exploration of optimal solutions without significantly increasing circuit depth, making it suitable for near-term applications. We test this approach on a portfolio optimization problem using an IBM 127 qubits Eagle processor (\texttt{ibm\_quebec}) and simulators. Compared to state-of-the-art algorithms, our results show that the proposed method improves fitness values by 33.6\% over classical genetic algorithm and 37.2\% over quantum-inspired genetic algorithm, using the same iteration counts and population sizes with real quantum hardware employing 100 qubits. These findings highlight the potential of current quantum computers to address real-world utility-scale combinatorial optimization problems.
\end{abstract}


\section{Introduction}
Optimization problems are prevalent in many fields, including engineering and finance. One type of computationally intensive problem is combinatorial optimization, where the goal is to select some objects from a set to maximize an objective function \cite{wolsey2014integer}. The knapsack problem is a well-known example of this, while a practical example is portfolio optimization, where the objective is to choose assets from a given set to maximize returns.
\par
As the size of the problem increases, the number of possibilities grows exponentially, making it classically intractable to find the exact solution\cite{hromkovivc2013algorithmics}. This is where alternative algorithms come into play to approximate the solution. Methods such as evolutionary algorithms and quantum computing are employed to address these challenges. While these algorithms do not always guarantee finding the optimal solution, they use heuristics to significantly speed up the process \cite{elbeltagi2005comparison, Bub_2010, farhi2014quantum}. 
\par
Genetic Algorithm (GA) are one of the primary methods used to solve combinatorial problems on an industrial scale \cite{eiben2015introduction, telikani2021evolutionary}. 
\par
On the other hand, quantum computing offers a potential solution to computationally hard optimization problems by leveraging quantum mechanics concepts like superposition and entanglement to perform computations at speeds beyond what classical computing can achieve \cite{Bub_2010, nielsen2010quantum}.
\par
By combining these two powerful methods—GA and quantum computing—it is possible to leverage the strengths of both simultaneously. The result is the Quantum Genetic Algorithm (QGA), in which quantum circuits are used to generate solutions to optimization problems \cite{lahoz2016quantum}. The genetic algorithm handles updating circuit parameters or generating new circuits in each generation. By measuring the circuits at each iteration, binary outputs are produced, which is the desired format for combinatorial optimization.
\par
Several QGAs have already been proposed \cite{lahoz2016quantum, wang2013improvement, ballinas2023hybrid, xiong2018quantum}. However, some are Quantum-Inspired Genetic Algorithms (QiGA), which are inspired by quantum mechanics but still perform computations classically, and some of them are unimplementable on actual quantum computers \cite{lahoz2016quantum}.
\newline
Although some variations of QiGA implementable on quantum computers have been proposed \cite{rubio2021quantum}, they do not leverage the entanglement. On the other hand, algorithms such as Reduced Quantum Genetic Algorithm (RQGA) \cite{udrescu2006implementing} or those proposed in \cite{acampora2022using} and \cite{ballinas2023hybrid} use amplitude amplification, or full entanglement, which require deep circuits. These circuits are not suitable for the Noisy Intermediate-Scale Quantum (NISQ) era \cite{preskill2018quantum} due to their circuit depth as deeper quantum circuits generally suffer from higher noise levels \cite{yanakiev2024dynamic, gaur2023noise} and the two-qubit gates are among the noisiest operations in current quantum computers \cite{ahsan2022quantum}.
\par
In this work, we propose a novel QGA that outperforms both classical GA and existing QiGA implementations. Our algorithm incorporates superposition and a limited number of entanglements, carefully managing two-qubit gates to keep the circuit depth minimal. The primary contribution of this work lies in the crossover step, where we introduce a novel method to find and encode the entanglement patterns for generating circuits.
\par
This crossover is designed to replicate the best solutions found in each generation by identifying correlation patterns and encoding them into circuits for the next generation. We use an elitism pool that includes the best results found up to each point in the algorithm. Using these solutions, we generate circuits by analyzing pairs of variables that appear to be correlated. For such bit pairs, we apply a CNOT gate, while for the rest, we use gates that increase the probability of measuring values that match the best solutions. This crossover mechanism enhances the probability of generating superpositions of high-quality solutions while encoding correlated variables within entangled states.
\par
This algorithm is applicable to a wide range of combinatorial optimization problems. In this paper, we evaluate its performance on a dataset for portfolio optimization. Experiments conducted on both simulators and superconducting IBM Quantum System One (\texttt{ibm\_quebec}) demonstrate that our algorithm outperforms classical GA and QiGA. Notably, on a subset of 100 assets, our algorithm surpasses the performance of classical GA on actual quantum hardware, underscoring its potential even in the near term. Moreover, by limiting the entanglements in the circuits, we ensure that the quantum circuit remains shallow, making the algorithm both suitable for the NISQ era \cite{gaur2023noise, yanakiev2024dynamic} and computationally efficient on simulators \cite{vidal2003efficient}.
\par
The rest of the paper is organized as follows: Section \ref{sec:background} provides the necessary background on GA, and QGA. Section \ref{sec:methodology} details the methodology of our proposed algorithm. The experimental results, obtained using both simulators and actual quantum computers, are presented in Section \ref{sec:expriments}. Finally, Section \ref{sec:conclusion} concludes the paper.

\section{Background}\label{sec:background}
\subsection{Genetic Algorithm}
Evolutionary algorithms are optimization methods inspired by natural selection, with genetic algorithms being a prominent example \cite{holland1992genetic}. In genetic algorithm, multiple random samples are generated, where each sample represents a solution made up of various variables. Each variable is called a gene, and these samples are referred to as chromosomes. The collection of all chromosomes at a given time (iteration) is known as a generation.
\par
Based on an optimization function, the best chromosomes are retained while the rest are eliminated. Using the top chromosomes from each generation, new chromosomes are created through a process called crossover. To increase diversity, mutations can be introduced, where there is a probability of randomly changing a gene within a chromosome.
\par
Consequently, the new generation for the next iteration includes the best samples from the current generation, new chromosomes produced from the offspring of solutions. This iterative process continues, producing successive generations with increasingly optimized chromosomes.
\par
There are various methods to select chromosomes for crossover, also known as parents \cite{razali2011genetic}. A common method is roulette wheel selection, in which each chromosome is chosen with a probability proportional to its fitness \cite{razali2011genetic}. Multiple approaches exist for performing crossover as well. For instance, in single-point crossover, a point is selected within the chromosome, where the first part of the offspring inherits genes from one parent (before the point), and the second part inherits genes from the other parent (after the point).
\par
For mutation, different methods can be applied. Examples include flipping a gene randomly from 0 to 1 or from 1 to 0 or swapping two genes within a chromosome.

\subsection{Quantum Genetic Algorithms}\label{sec:qiga}
QGA is the result of combining GA with quantum computing \cite{han2000genetic}. Unlike GA, where chromosomes are represented as binary strings that undergo crossover or mutation to generate a new population, in QGA, the initial genes are in the form of qubits. Binary chromosomes (solutions) are obtained by measuring these qubits. To generate the states of these qubits, quantum circuits are utilized, and by applying certain quantum gates or modifying circuit parameters (e.g., rotation angles), new circuits are created resulting in different solutions \cite{lahoz2016quantum}.
\par
Several QiGA variants and enhancements have been proposed \cite{xiong2018quantum}. QiGA's main idea is to encode genes in qubit-inspired format as below:

\[
|\psi\rangle = 
\begin{bmatrix}
\alpha_1 & \alpha_2 & \dots & \alpha_n \\
\beta_1  & \beta_2  & \dots & \beta_n
\end{bmatrix}
\]

Each qubit has two parameters of $\alpha$ and $\beta$ corresponding to the probabilities of being 0 and 1, respectively. Quantum chromosomes are generated using quantum circuits. The circuit is initialized in the zero state and then put into superposition using Hadamard gates. Rotation gates are applied to each qubit, and the final binary output is obtained by measuring the circuits.
\par
In each generation, based on a comparison of the best solution ($x^b$) and the solution obtained by each circuit ($x^{i}$), the rotation parameters are updated. As proposed in \cite{wang2013improvement}, the rotation direction for gene $j$ of the chromosome $i$ is determined by calculating a determinant using the probability amplitude of that gene ($\alpha^{i}_{j}$, and $\beta^{i}_{j}$) and the corresponding values of the best solution ($\alpha^{b}_{j}$, and $\beta^{b}_{j}$) as \eqref{eq:determinant}. The direction = $+1$, and direction = $-1$ means the clockwise, and counter clockwise rotations respectively.

\begin{equation}
      \text{direction} =
  \begin{cases}
    -\text{sgn}(D) & \text{if } D \neq 0, \\
    \pm 1 & \text{if } D = 0,
  \end{cases} \hspace{0.2cm} D = \det
\begin{vmatrix}
\alpha^{b}_{j} & \alpha^{i}_{j} \\
\beta^{b}_{j} & \beta^{i}_{j}
\end{vmatrix}
\label{eq:determinant}
\end{equation}

An adaptive method is used to determine the magnitude of the rotation adjustment in self-adaptive QGA (AQGA)\cite{wang2013improvement}: 

\begin{equation}
    \theta_i = \theta_{\text{max}} - \frac{\theta_{\text{max}} - \theta_{\text{min}}}{\text{iter}_{\text{max}}} \times \text{iter}
    \label{eq:amplitude}
\end{equation}
where $\theta_{max}$, and $\theta_{min}$ are hyperparameters defining the rotation angles, $iter$ is the iteration number, and $iter_{max}$ is the maximum number of iterations.
\par
Some heuristics, such as adding an $X$ gate on random positions to introduce mutations, are also incorporated into the algorithm. However, since updating the rotation angles requires access to $\alpha$ and $\beta$, AQGA cannot be implemented on actual quantum computers.
\par
Several QGAs that are implementable on actual quantum computers have been proposed as well \cite{acampora2022using, rubio2021quantum, ballinas2023hybrid}. In \cite{rubio2021quantum}, a new rotation scheme is introduced by utilizing the expectation value of each qubit, making it feasible for actual quantum hardware. In \cite{montiel2019quantum}, a quantum evolutionary algorithm inspired by the \textit{Acromyrmex} ant species is proposed. Additionally, the RQGA, leveraging Grover's algorithm, was introduced \cite{udrescu2006implementing}. However, due to the large circuit depth required by Grover's algorithm, RQGA is mostly suitable for fault-tolerant quantum computing (FTQC) \cite{zhang2021implementation}.

\subsection{Portfolio Optimization}\label{sec:PO}
In finance, a portfolio is a collection of assets, and the goal of portfolio optimization is to maximize returns from these assets while managing risk \cite{buonaiuto2023best}. Among the various approaches to portfolio optimization, the Mean-variance (MinVar) formulation \cite{markowitz2000mean} is one of the most widely used \cite{gunjan2023brief}. Based on modern portfolio theory introduced by Markowitz \cite{markowitz1952modern}, it aims to maximize returns while minimizing risk, where risk is typically defined as the variance of the returns of the portfolio \cite{buonaiuto2023best}.
\par
For a portfolio of multiple assets, we define $x$ as the asset allocation vector, where $x_i = 1$ indicates the inclusion of asset $i$ in the portfolio, and $x_i = 0$ indicates its exclusion. Let $\mu$ represent the vector of mean returns for the assets, and $\Sigma$ represent the covariance matrix of asset returns. 
\par
The portfolio optimization problem can be formulated as a Quadratic Unconstrained Binary Optimization (QUBO) problem, as shown in \eqref{eq:PO} \cite{buonaiuto2023best}. 

\begin{equation}
\begin{split}
\max_x \ \mu^T x - q \cdot x^T \Sigma x 
\\
\text{subject to:} \quad
x_i \in \{0, 1\} \quad \forall i .
\end{split}\label{eq:PO}
\end{equation}

In this formulation, $q$ represents the risk aversion coefficient, where a higher $q$ signifies lower tolerance for risk \cite{buonaiuto2023best}.
\newline
Return ($\mu$) and risk ($\Sigma$) are calculated using historical data over a specific time period.
\par
Additional constraints may include limits on the number of assets in the portfolio or defining $x$ as continuous variables between 0 and 1, representing the proportion of each asset in the portfolio. However, in this work, we focus on the binary case where $x$ is either 0 or 1.
\par
This problem is NP-hard \cite{cesarone2015linear}, and as the number of assets (dimensions of $x$) increases, traditional solvers require a prohibitively long time to solve it. Therefore, heuristic methods and evolutionary algorithms are often used as alternative approaches \cite{erwin2023meta, gunjan2024quantum}.

\section{Methodology}\label{sec:methodology}
This section presents our proposed QGA, which generates quantum circuits for the next generation based on measurement results from the current one.
\par
We maintain an elitism pool \( Pool_{\text{elitism}} = \{ x^{b_1}, x^{b_2} \mid x_i \in \{0, 1\}^n \} \), containing the top two solutions with the highest fitness values at each iteration. The pool is updated and sorted by fitness after each generation. To evaluate and rank solutions, we use the objective function defined in Equation~\ref{eq:PO} as the fitness metric, where higher values indicate better solution quality. 

Each iteration involves generating quantum circuits and measuring them once to produce binary solutions, which are then associated with their respective circuits.

\begin{figure}
    \centering
    \includegraphics[width=0.48\textwidth]{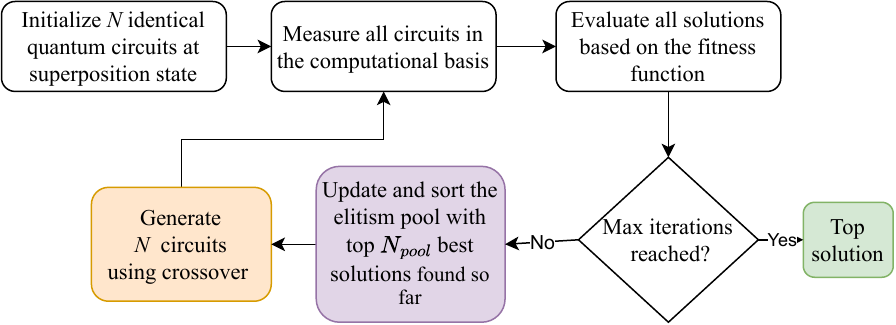}
    \caption{Diagram of our proposed QGA.}
    \label{fig:diagram}
\end{figure}

\par
Fig. \ref{fig:diagram} illustrates the block diagram of our algorithm. In the following sections, we describe each step of the algorithm in more detail.

\subsection{Initialization}
For initialization, we apply Hadamard gates to each qubit, creating a superposition of all computational basis states. This process is analogous to the initial Hamiltonian in quantum annealing, allowing us to explore a wide range of potential solutions \cite{annealing-phase}. We then create \( N \) copies of this circuit (where \( N \) represents the population size) and measure them once in the computational basis. To accomplish this, we utilize the Qiskit Sampler primitive \cite{QiskitPrimitives}, resulting in \( N \) chromosomes, each composed of binary genes.

\subsection{Crossover}
In the crossover step, we propose a heuristic approach aimed at generating circuits that yield the best solutions (which produced higher fitness values) or combinations of those solutions. Initially, the solutions were generated randomly, as every qubit was in the \( |+\rangle \) state in all circuits and randomly measured as 0 or 1 with equal probability.  Our objective is to construct new quantum circuits that increase the likelihood of sampling bitstrings corresponding to high-quality solutions.

Two parent solutions, \( x^{b_1} \) and \( x^{b_2} \), from the elitism pool are used to generate \( N \) new circuits for the next generation.

The core idea is to exploit entanglement patterns present in high-quality solutions. Specifically, we identify pairs of bits that exhibit correlation across the two parents. These correlated bit pairs are then mapped to entangled qubit pairs in the next generation’s circuit.

To detect these correlations, we search for Bell-state-like patterns, characterized by bit pairs with either matching values (\( |00\rangle \) and \( |11\rangle \)) or opposing values (\( |01\rangle \) and \( |10\rangle \)) across both parents. 

\par
\begin{figure}
    \centering
    \includegraphics[width=0.2\textwidth]{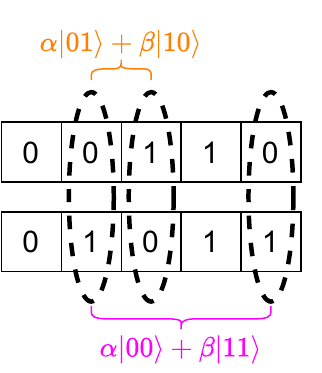}
    \caption{Entanglement traits in two solutions. We can consider qubits 1 and 2 of both solutions as the \( |\psi^n\rangle = \alpha|01\rangle + \beta|10\rangle \) state and qubits 1 and 4 as the\(|\psi^p\rangle = \alpha|00\rangle + \beta|11\rangle \) state.}
    \label{fig:cross2_bits}
\end{figure}

For example, consider a problem with five variables (zero-indexed) where the two parent solutions are \( x^{b_1} = [0,\, 0,\, 1,\, 1,\, 0] \) and \( x^{b_2} = [0,\, 1,\, 0,\, 1,\, 1] \). In this case, bits 1 and 2 have opposite values across the two parents, while bits 1 and 4 have matching values (see Fig.~\ref{fig:cross2_bits}). 

We denote the entangled state \( \alpha|00\rangle + \beta|11\rangle \) as \( |\psi^p\rangle \), representing a positive correlation, and the entangled state \( \alpha|01\rangle + \beta|10\rangle \) as \( |\psi^n\rangle \), representing a negative correlation.

All such bit pairs are considered candidates for entanglement. However, not all are selected. Instead, we apply a probabilistic selection strategy that takes into account their corresponding coupling coefficients in the QUBO formulation.

Let $\Sigma \in \mathbb{R}^{n \times n}$ be a covariance matrix, and let $\Sigma_n$ be its normalized form, defined as:
\[
\Sigma_n = \frac{\Sigma}{\max_{i,j} |\Sigma_{i,j}|}
\]

Define two sets:
\begin{itemize}
    \item $\mathcal{P}$: the set of index pairs $(i,j)$ with positive correlation detected from binary results 
    \item $\mathcal{N}$: the set of index pairs $(i,j)$ with negative correlation detected from binary results
\end{itemize}

Let $p_s \in \mathbb{R}$ be a hyperparameter controlling the pair selection probability, and define the decay factor ($df$) at iteration $t$ as:
\[
df(t) = 0.5 + \frac{t}{2 \cdot T_{\max}}
\]
where $T_{\max}$ is the maximum number of iterations.

Then, for each pair $(i,j)$, the pair selection probability $p_s^{(i,j)}$ is defined as follows:

\[
p_s^{(i,j)} = 
\begin{cases}
p_s \cdot df(t) \cdot |\Sigma_n^{(i,j)}|, & \text{if } (i,j) \in \mathcal{P} \text{ and } \Sigma_n^{(i,j)} > 0 \\
p_s \cdot |\Sigma_n^{(i,j)}|, & \text{if } (i,j) \in \mathcal{P} \text{ and } \Sigma_n^{(i,j)} \leq 0 \\
p_s \cdot |\Sigma_n^{(i,j)}|, & \text{if } (i,j) \in \mathcal{N} \text{ and } \Sigma_n^{(i,j)} < 0 \\
p_s \cdot df(t) \cdot |\Sigma_n^{(i,j)}|, & \text{if } (i,j) \in \mathcal{N} \text{ and } \Sigma_n^{(i,j)} \geq 0 \\
\end{cases}
\]

In this approach, the influence of the coupling coefficient is modulated over the course of the optimization process. Early in the iterations, if a positively (or negatively) correlated pair $(i, j)$ does not align with the sign of the corresponding coupling coefficient $\Sigma^{(i,j)}$, its selection probability is penalized by a decay factor starting at 0.5. This reduces the likelihood of selecting pairs that contradict the objective function's structure at early stages, enforcing consistency with the QUBO formulation.
\par
As the algorithm progresses, the decay factor gradually increases toward 1.0, removing this penalty. By the final iteration, all candidate pairs are treated uniformly in terms of selection probability magnitude, regardless of initial alignment with the coupling coefficients. This reflects a shift in emphasis from structural conformity to empirical guidance, allowing the optimization to favor pairs supported by iterative evidence rather than initial coupling assumptions.   
\par
After identifying all correlated qubit pairs, we examine them for shared qubits to form entanglement chains. For example, if qubit $q_1$ is entangled with qubit $q_2$, and qubit $q_2$ with qubit $q_4$, we construct a chain involving qubits $q_1$, $q_2$, and $q_4$. One qubit in the chain is selected as the control, while the others are designated as targets, enabling multi-qubit entanglement through chained operations. This ensures that the number of entangled pairs is at most one less than the number of qubits.

\begin{figure}
    \centering
    \subfigure[Quantum circuit for \( |\psi^p\rangle=\alpha|00\rangle + \beta|11\rangle \)]{
        \includegraphics[width=0.22\textwidth]{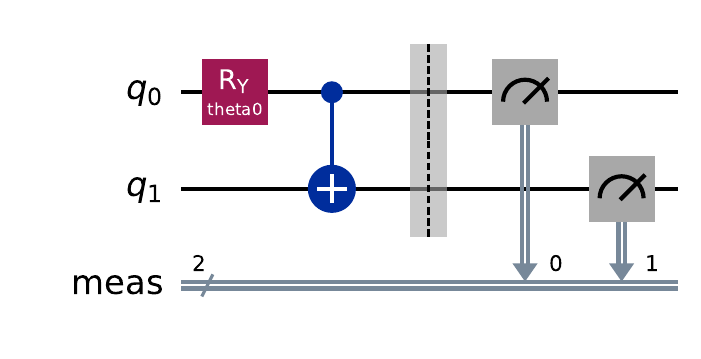}
        \label{fig:cross1_a}
    }
    \hspace{1mm}
    \subfigure[Quantum circuit for \( |\psi^n\rangle=\alpha|01\rangle + \beta|10\rangle \)]{
        \includegraphics[width=0.22\textwidth]{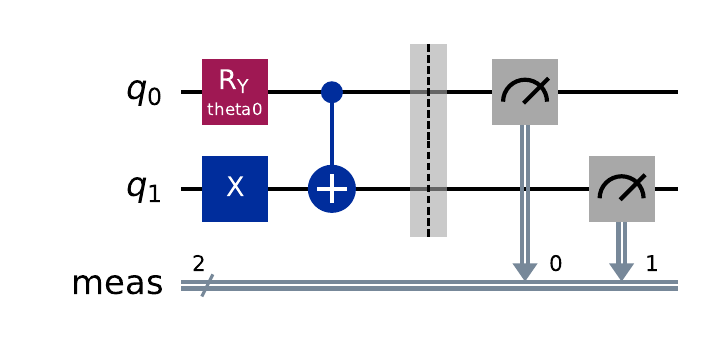}
        \label{fig:cross1_b}
    }
    \caption{Two possible circuits to form the entanglement states. Rotation with the angle of $\theta$ controls the amplitude of the states ($\alpha$ and $\beta$)}
    \label{fig:cross1}
\end{figure}

Once all entangled pairs and chains are identified, we encode them into the quantum circuits for the next generation. This is achieved using a combination of CNOT gates and rotation gates, as illustrated in Fig. \ref{fig:cross1}. Fig. \ref{fig:cross1_a} shows the circuit used to create positive correlation ($|\psi^p\rangle$), while Fig. \ref{fig:cross1_b} illustrates the circuit for generating negative correlation ($|\psi^n\rangle$). The CNOT gates establish entanglement between control and target qubits, while the rotation gates applied to the control qubits allow us to manipulate the probability amplitudes ($\alpha$ and $\beta$) of the resulting quantum states.

Our goal is to configure the circuit such that measurements of the entangled qubit pairs yield outcomes reflecting the desired bit correlations. These correlations correspond to superpositions similar to entangled states, with outcome probabilities weighted in favor of the bit values observed in the best solution. The rotation angle for the control qubit is to control this probability (\( p_a \)).

Through this approach, we ensure that highly correlated qubits remain entangled across generations, preserving their joint structure while allowing for variation in their specific values. This maintains the beneficial structural patterns discovered in earlier solutions while promoting controlled exploration. 
\par
For bits that do not exhibit an entanglement pattern with any other bit, we apply single-qubit rotation gates to control the probability distribution of their measurement outcomes. Specifically, we use the same hyperparameter \( p_a \) to define the probability of measuring the bit in the value corresponding to the best solution.

If the bit value in the best solution is \( 0 \), we aim for the qubit to be measured in the \( |0\rangle \) state with probability \( p_a \). To achieve this, we apply an \( RY \) rotation gate with angle:

\[
\theta = 2 \cdot \arccos\left(\sqrt{p_a}\right)
\]

Conversely, if the desired measurement outcome is \( 1 \), we want the qubit to be measured in the \( |1\rangle \) state with probability \( p_a \), and the corresponding angle is set as:

\[
\theta = 2 \cdot \arccos\left(\sqrt{1 - p_a}\right)
\]

This method ensures that even non-entangled qubits are biased toward values found in the best solutions, guiding the circuit generation toward favorable regions of the search space while maintaining probabilistic diversity.
\par
\begin{figure}
    \centering
    \includegraphics[width=0.48\textwidth]{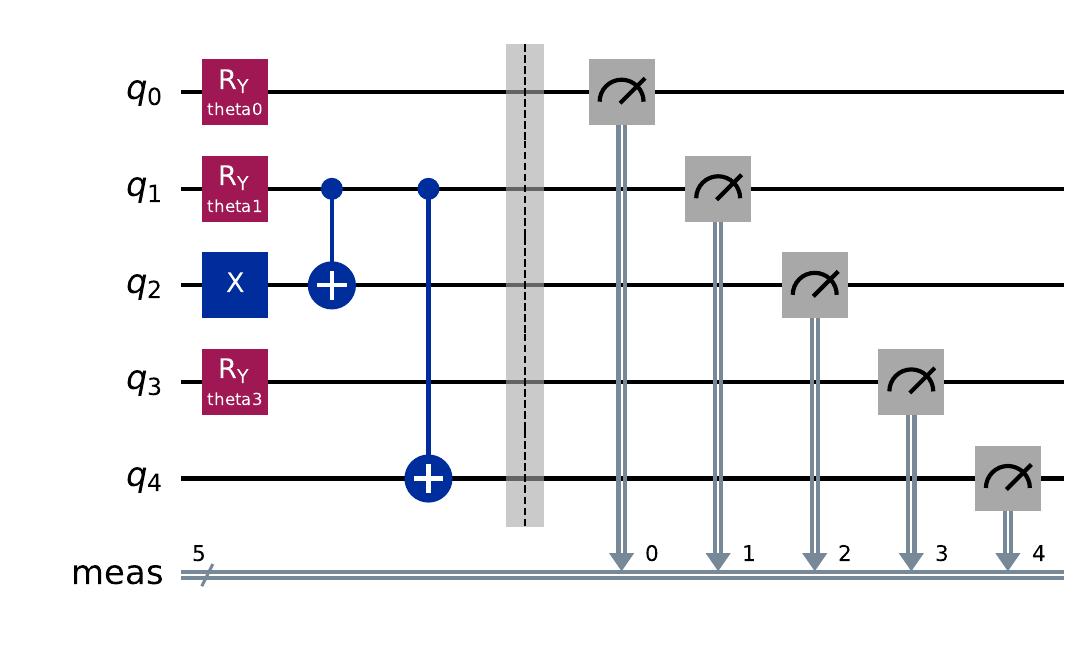}
    \caption{Corresponding circuit using the parents from Fig. \ref{fig:cross2_bits}.}
    \label{fig:cross2}
\end{figure}

Fig. \ref{fig:cross2} illustrates a possible quantum circuit derived from the bitstrings shown in Fig. \ref{fig:cross2_bits}. Note that in this example, qubits 2 and 4 also exhibit entanglement patterns. However, under the probabilistic selection scheme for entangled pairs, only the pairs (1,~2) and (1,~4) are selected for entanglement in this particular instance. Different subsets of entangled pairs may be selected in other circuits generated within the same generation, resulting in variations in circuit structure across the population.

Generating multiple circuits based on this solution leads to variations where specific combinations of entanglements are selected. This enables us to explore different entangled possibilities and detect those that are correlated.
\par
This crossover mechanism is analogous to classical crossover and mutation, where each bit in the offspring is inherited from one of the parent solutions, with the possibility of mutation.  

However, unlike classical methods, this approach also encodes entanglement patterns into the offspring. In other words, if one bit from a correlated pair is selected for crossover or mutation, its entangled counterpart is also subjected to the same operation.

Each circuit is measured, and the elitism pool is updated and sorted accordingly. Circuits for the next generation are then created based on the updated elitism pool. This process repeats until a stopping criterion is met.

Algorithm 1 describes the pseudocode for this algorithm. \( S \) is the number of bits in a solution. \( QC^{c}(n) \) is the \( n^{\text{th}} \) qubit of the circuit \( QC^{c} \) to be generated by crossover, and \( x^{b_1}_n \), \( x^{b_2}_n \) are the \( n^{\text{th}} \) bits of the parents. The set \texttt{SelectedEntangledQubits} denotes the collection of all qubit pairs that have been selected for entanglement in a given circuit.

\let\oldnl\nl% Store \nl in \oldnl
\newcommand{\nonl}{\renewcommand{\nl}{\let\nl\oldnl}}
\begin{algorithm}
\caption{EAQGA}
\KwData{QUBO matrix: $Q$, Population size $N$, Maximum iterations $T_{max}$, Selection and Amplitude probabilities $p_s$ and $p_a$}
\KwResult{Best solution $x^{b_1}$}

\textbf{Initialization:}\\
Initialize a quantum circuit $QC_0$ in the state $\psi_0 = |0\rangle^{\otimes S}$\;
\For{Population $p = 1$ to $N$}{
    $QC^{p} \gets QC_0$\;
    \For{Qubit $n = 0$ to $S-1$}{
        $QC^{p}.H(n)$\; 
    }
}
\For{Iteration $i = 1$ to $T_{max}$}{
\textbf{Evaluation:}\\
\For{Population $p = 1$ to $N$}{
    Measure circuit $QC^p$ in the computational basis to obtain $x^p$\;
}
\textbf{Crossover:}\\
Update the elitism pool: $Pool_{elitism} \gets \{x^{b_1}, x^{b_2}\}$\;
Initialize a quantum circuit $QC_0$ in the state $\psi_0 = |0\rangle^{\otimes n}$\;
\For{Population $p = 1$ to $N$}{
    Select the entangled qubits using $Pool_{elitism}$, $P_{s}$, and $Q$:
    $SelectedEntangledQubits$\;
    $QC^{p} \gets QC_0$\;
    \For{$(n_1, n_2) \in SelectedEntangledQubits$}{
        \If{$(x^{b_1}_{n_1} == 0$}{
        $QC^{p}.RY(2* \arccos(\sqrt{P_a}), n_1)$\;
        $QC^{p}.CX(n_1, n_2)$\;
        }
        \Else{
        $QC^{p}.RY(2* \arccos(\sqrt{1-P_a}), n_1)$\;
        $QC^{p}.X(n_2)$\;
        $QC^{p}.CX(n_1, n_2)$\;
        }
    }
    
    \For{for $n = 1$ to $S$}{
        \If{$n_1 \notin SelectedEntangledQubits $}{
                \If{$x^{b_1}_{n_1} == 0$}{
                    $\theta = 2* \arccos(\sqrt{P_a})$\;
                }
                \Else{
                    $\theta = 1 - 2* \arccos(\sqrt{P_a})$\;
                }
                $QC^{p}.RY(\theta, n)$\;
            }
        }
    }
}

\textbf{Return:} $x^{b_1}$ 
\end{algorithm}

\section{Experiments}\label{sec:expriments}
In this section, we present the experimental results of our algorithm and compare them with those of GA and AQGA. While our proposed algorithm is designed to be problem-agnostic and applicable to any combinatorial optimization task, we focused our experiments on the practical problem of portfolio optimization. These experiments were conducted using both simulators and an actual quantum computer, with subsets of varying sizes. Below, we provide a detailed overview of the dataset, algorithms, and numerical results.

\subsection{Algorithms}
We included three algorithms in our experiments: GA, AQGA, and our proposed algorithm. The hyperparameters for all three algorithms were optimized using Bayesian optimization \cite{bayesian-optimize} on a randomly selected subset from our dataset.

\subsubsection{GA}
The initial population was generated using random binary bitstrings. Parent selection was performed via fitness-proportional (roulette wheel) selection, with fitness values shifted to be non-negative. Single-point crossover was applied with a probability of 0.85, and mutation was performed by flipping a random bit with a rate of 0.03.

\subsubsection{AQGA}
For the QiGA, we used the adaptive version (AQGA) proposed in \cite{wang2013improvement} and summarized in Section \ref{sec:qiga}. In this algorithm, the direction of rotation angles is calculated by the sign of the determinant in \eqref{eq:determinant}. For calculating the amplitude, we used \eqref{eq:amplitude}, setting \( \theta_{\text{max}} = 0.25 \) and \( \theta_{\text{min}} = 0.15 \). The mutation is done by flipping the \( \alpha \) and \( \beta \) of a random gene in a chromosome. The mutation ratio was set to 0.05. 
\newline
Moreover, as proposed in \cite{wang2013improvement}, a new operation called the disaster operation, introduced to improve performance, was included in our implementation. The idea behind the disaster operation is to check if the solutions have not improved for a specific number of iterations (6 in our experiments). If true, a proportion of the worst solutions (0.2 in our experiment) would be reinitialized in equal superposition again, i.e.,

\[
|\psi\rangle = 
\begin{bmatrix}
\alpha_1 & \alpha_2 & \dots & \alpha_n \\
\beta_1  & \beta_2  & \dots & \beta_n
\end{bmatrix}
=
\begin{bmatrix}
\frac{1}{\sqrt{2}} & \frac{1}{\sqrt{2}} & \dots & \frac{1}{\sqrt{2}} \\
\frac{1}{\sqrt{2}} & \frac{1}{\sqrt{2}} & \dots & \frac{1}{\sqrt{2}}
\end{bmatrix}
\]
for those chromosomes selected by the disaster operator.

\subsubsection{EAQGA}
Since our proposed algorithm considers entanglement patterns in the crossover process, we named it the Entanglement-Aware Quantum-Enhanced Genetic Algorithm (EAQGA).
\par
The following hyperparameters were used for our algorithm: 
\begin{itemize}
    \item In each circuits, the probability of reproducing the bits from the best solution is  ($p_{a} = 0.95$).
    \item The selection probability for bit pairs is set to ($p_s = 0.6$).
\end{itemize}
We measured each circuit only once to produce the solution.

\subsection{Dataset}
We experimented with real stocks of companies derived from the S\&P 500 index, which includes 500 of the largest publicly traded companies in the United States. This index covers approximately 80\% of the total market capitalization of U.S. public companies \cite{sp5002024}. We used historical data from 01/10/2023 to 30/09/2024 with a time interval of one day. The DataProvider class of Qiskit Finance was used to obtain the data \cite{qiskit_stock_data}. Specifically, we utilized YahooDataProvider \cite{qiskit_stock_data} in Qiskit to retrieve data of the S\&P 500 index from Yahoo Finance \cite{yahoo_finance}.
\par
As the objective function, the MinVar formulation introduced in Section \ref{sec:PO} was utilized. We set the risk aversion coefficient (\( q \)) equal to 0.5. To calculate the average (\( \mu \)) and covariance (\( \Sigma \)), historical data were used with the following formulation, where \( P_{it} \) is the price of asset \( i \) at time \( t \). $n$ is the number of assets, and $T$ is the time length \cite{buonaiuto2023best}.

\begin{equation}
\begin{split}
\mu_i = \frac{1}{T} \sum_{t=1}^{T} R_{it}, \quad \Sigma_{i, j} = \frac{1}{T-1} \sum_{t=1}^{T} (R_{it} - \mu_i)(R_{jt} - \mu_j) \\
\text{where } R_{it} = \frac{P_{it}}{P_{it-1}} - 1 \quad \text{for } i = 1, 2, \ldots, n
\end{split}
\end{equation}

To generate subsets of specific sizes, we randomly chose sets of a specific number of stocks from the S\&P 500. For experiments on the simulator, we generated 10 subsets of 30 stocks and 10 subsets of 40 stocks. subsets of the same length did not overlap in stocks. For experiments on the real quantum computer, we used a subset of 100 randomly chosen assets.

\subsection{Computation Resource}
For the experiments in Section \ref{sec:simulator}, we used the Qiskit Matrix Product State (MPS) simulator \cite{qiskit_mps_tutorial}. MPS is a tensor-network statevector simulator that represents the quantum state using a Matrix Product State \cite{schollwock2011density, vidal2003efficient}. This simulator is efficient for circuits with relatively low entanglements \cite{vidal2003efficient}, making it particularly suitable for our algorithm.
\par
In Section \ref{sec:real}, we conducted experiments on an actual quantum computer, specifically IBM's 127-qubit \texttt{ibm\_quebec} system \cite{IMB_PINQ}.

\subsection{Results}
\subsubsection{On Simulator}\label{sec:simulator}
The algorithms were run on 10 subsets of size 30 and 10 subsets of size 40. For all algorithms, we used 20 iterations. We performed experiments with population sizes of 10 and 20. Each algorithm was run 100 times on each subset, and the average and standard deviation of the fitness values were calculated. Table \ref{tab:results 30} presents the results for subsets of size 30, while Table \ref{tab:results 40} shows the results for subsets of size 40. The maximum average values across the three algorithms for each population size and subset are highlighted in bold. Additionally, we included the optimal (maximum) result obtained using brute-force search.

\renewcommand{\arraystretch}{1}
\begin{table*}[t]
\centering
\caption{Fitness value of the three algorithms with a problem size of 30 and population sizes of 10 and 20 on the simulator over 20 iterations. All values were multiplied by 100 for improved readability.}
\setlength{\tabcolsep}{5pt}
\fbox{ 
\begin{tabular}{c|c|cc@{\hspace{18pt}}cc@{\hspace{18pt}}cc|cc@{\hspace{18pt}}cc@{\hspace{18pt}}cc}
\multirow{2}{*}{\shortstack{Subset}} & \multirow{2}{*}{Optimum} & \multicolumn{6}{c|}{Population = 10} & \multicolumn{6}{c}{Population = 20} \\
\cmidrule(lr){3-8} \cmidrule(lr){9-14}
 & & \multicolumn{2}{c}{\textbf{GA \phantom{B}}} & \multicolumn{2}{c}{\textbf{AQGA \phantom{BBB}}} & \multicolumn{2}{c|}{\textbf{EAQGA \phantom{B}}} & \multicolumn{2}{c}{\textbf{GA \phantom{B}}} & \multicolumn{2}{c}{\textbf{AQGA \phantom{BBB}}} & \multicolumn{2}{c}{\textbf{EAQGA \phantom{BBB}}} \\
\cmidrule(lr){3-14}
 & & Avg & Std & Avg & Std & Avg & Std & Avg & Std & Avg & Std & Avg & Std \\
\midrule
     1 & 2.0302 & 1.7630 & 0.1049 & 1.8422 & 0.0805 & \textbf{1.9946} & 0.0390 & 1.8711 & 0.0760 & 1.9322 & 0.0536 & \textbf{2.0258} & 0.0111 \\
     2 & 2.3011 & 1.9865 & 0.1131 & 2.0716 & 0.0959 & \textbf{2.2642} & 0.0471 & 2.1140 & 0.0860 & 2.1988 & 0.0600 & \textbf{2.2961} & 0.0171 \\
     3 & 2.0612 & 1.8245 & 0.0869 & 1.9092 & 0.0515 & \textbf{2.0368} & 0.0254 & 1.9152 & 0.0603 & 1.9937 & 0.0363 & \textbf{2.0568} & 0.0082 \\
     4 & 2.1863 & 1.9741 & 0.0875 & 2.0573 & 0.0570 & \textbf{2.1638} & 0.0245 & 2.0437 & 0.0606 & 2.1208 & 0.0318 & \textbf{2.1825} & 0.0073 \\
     5 & 2.2306 & 1.9827 & 0.0982 & 2.0770 & 0.0663 & \textbf{2.1996} & 0.0259 & 2.0792 & 0.0646 & 2.1531 & 0.0411 & \textbf{2.2246} & 0.0112 \\
     6 & 2.2754 & 2.0155 & 0.1112 & 2.1163 & 0.0688 & \textbf{2.2459} & 0.0276 & 2.1212 & 0.0736 & 2.2022 & 0.0376 & \textbf{2.2698} & 0.0136 \\
     7 & 1.8252 & 1.6076 & 0.0867 & 1.6853 & 0.0547 & \textbf{1.8003} & 0.0286 & 1.7002 & 0.0595 & 1.7584 & 0.0337 & \textbf{1.8208} & 0.0102 \\
     8 & 1.8903 & 1.7092 & 0.0740 & 1.7822 & 0.0522 & \textbf{1.8716} & 0.0190 & 1.7821 & 0.0469 & 1.8386 & 0.0286 & \textbf{1.8866} & 0.0064 \\
     9 & 2.2308 & 1.9616 & 0.1049 & 2.0489 & 0.0780 & \textbf{2.1925} & 0.0368 & 2.0418 & 0.0755 & 2.1426 & 0.0460 & \textbf{2.2265} & 0.0097 \\
    10 & 2.5445 & 2.3301 & 0.0901 & 2.4089 & 0.0590 & \textbf{2.5259} & 0.0216 & 2.4114 & 0.0626 & 2.4788 & 0.0360 & \textbf{2.5417} & 0.0070 \\
\midrule
    \textbf{Average} & 2.1576 & 1.9155 &  & 1.9999 &  & \textbf{2.1295} &  & 2.0080 &  & 2.0819 &  & \textbf{2.1531} &  \\

\end{tabular}
}
\setlength{\tabcolsep}{6pt}
\label{tab:results 30}
\end{table*}

\begin{table*}[t]
\centering
\caption{Fitness value of the three algorithms with a problem size of 40 and population sizes of 10 and 20 on the simulator over 20 iterations. All values were multiplied by 100 for improved readability.}
\setlength{\tabcolsep}{5pt}
\fbox{ 
\begin{tabular}{c|c|cc@{\hspace{18pt}}cc@{\hspace{18pt}}cc|cc@{\hspace{18pt}}cc@{\hspace{18pt}}cc}
\multirow{2}{*}{\shortstack{Subset}} & \multirow{2}{*}{Optimum} & \multicolumn{6}{c|}{Population = 10} & \multicolumn{6}{c}{Population = 20} \\
\cmidrule(lr){3-8} \cmidrule(lr){9-14}
 & & \multicolumn{2}{c}{\textbf{GA \phantom{B}}} & \multicolumn{2}{c}{\textbf{AQGA \phantom{BBB}}} & \multicolumn{2}{c|}{\textbf{EAQGA \phantom{B}}} & \multicolumn{2}{c}{\textbf{GA \phantom{B}}} & \multicolumn{2}{c}{\textbf{AQGA \phantom{BBB}}} & \multicolumn{2}{c}{\textbf{EAQGA \phantom{BBB}}} \\
\cmidrule(lr){3-14}
 & & Avg & Std & Avg & Std & Avg & Std & Avg & Std & Avg & Std & Avg & Std \\
\midrule
     11 & 2.7435 & 2.3518 & 0.1231 & 2.4544 & 0.0885 & \textbf{2.6637} & 0.0524 & 2.4649 & 0.0744 & 2.5617 & 0.0618 & \textbf{2.7169} & 0.0231 \\
     12 & 2.8257 & 2.4089 & 0.1249 & 2.5293 & 0.0949 & \textbf{2.7411} & 0.0562 & 2.5265 & 0.0902 & 2.6445 & 0.0690 & \textbf{2.8033} & 0.0221 \\
     13 & 2.7337 & 2.3148 & 0.1126 & 2.4131 & 0.0899 & \textbf{2.6394} & 0.0503 & 2.4287 & 0.0897 & 2.5298 & 0.0665 & \textbf{2.7087} & 0.0248 \\
     14 & 2.5040 & 2.1936 & 0.0915 & 2.2721 & 0.0560 & \textbf{2.4428} & 0.0354 & 2.2765 & 0.0721 & 2.3614 & 0.0460 & \textbf{2.4864} & 0.0169 \\
     15 & 2.5634 & 2.1647 & 0.1029 & 2.2859 & 0.0867 & \textbf{2.4901} & 0.0435 & 2.2888 & 0.0831 & 2.4011 & 0.0576 & \textbf{2.5448} & 0.0221 \\
     16 & 1.7853 & 1.4394 & 0.1136 & 1.5282 & 0.0865 & \textbf{1.7195} & 0.0416 & 1.5751 & 0.0737 & 1.6268 & 0.0539 & \textbf{1.7633} & 0.0213 \\
     17 & 2.0170 & 1.6372 & 0.1033 & 1.7409 & 0.0884 & \textbf{1.9407} & 0.0450 & 1.7555 & 0.0771 & 1.8374 & 0.0638 & \textbf{1.9910} & 0.0228 \\
     18 & 2.1908 & 1.7217 & 0.1584 & 1.8034 & 0.1120 & \textbf{2.0953} & 0.0697 & 1.8807 & 0.1127 & 1.9639 & 0.0819 & \textbf{2.1552} & 0.0318 \\
     19 & 2.8334 & 2.3876 & 0.1360 & 2.5075 & 0.0828 & \textbf{2.7321} & 0.0521 & 2.4949 & 0.1026 & 2.6209 & 0.0691 & \textbf{2.8076} & 0.0226 \\
     20 & 2.4792 & 2.0529 & 0.1371 & 2.1428 & 0.0992 & \textbf{2.3935} & 0.0601 & 2.1869 & 0.0893 & 2.2894 & 0.0751 & \textbf{2.4576} & 0.0217 \\
\midrule
    \textbf{Average} & 2.4676 & 2.0673 &  & 2.1678 &  & \textbf{2.3858} &  & 2.1879 &  & 2.2837 &  & \textbf{2.4435} &  \\
\end{tabular}
}
\setlength{\tabcolsep}{6pt}
\label{tab:results 40}
\end{table*}

The average results from 100 runs demonstrated that EAQGA achieved the best outcomes among the three algorithms across the dataset, regardless of subset size or population configuration. Notably, even with a smaller population size of 10, EAQGA surpassed GA and AQGA, both of which required a larger population size of 20 to approach comparable performance levels.
\par
In addition to achieving higher fitness values, EAQGA exhibited higher stability, as indicated by its consistently lower standard deviations. This robustness ensured reliable and predictable outcomes across repeated runs.
\par
For subsets with larger problem size (size 40), EAQGA maintained its performance advantage, effectively handling the increased complexity.
\par
\begin{figure}
    \centering
    \subfigure[Size 30]{
        \includegraphics[width=0.225\textwidth]{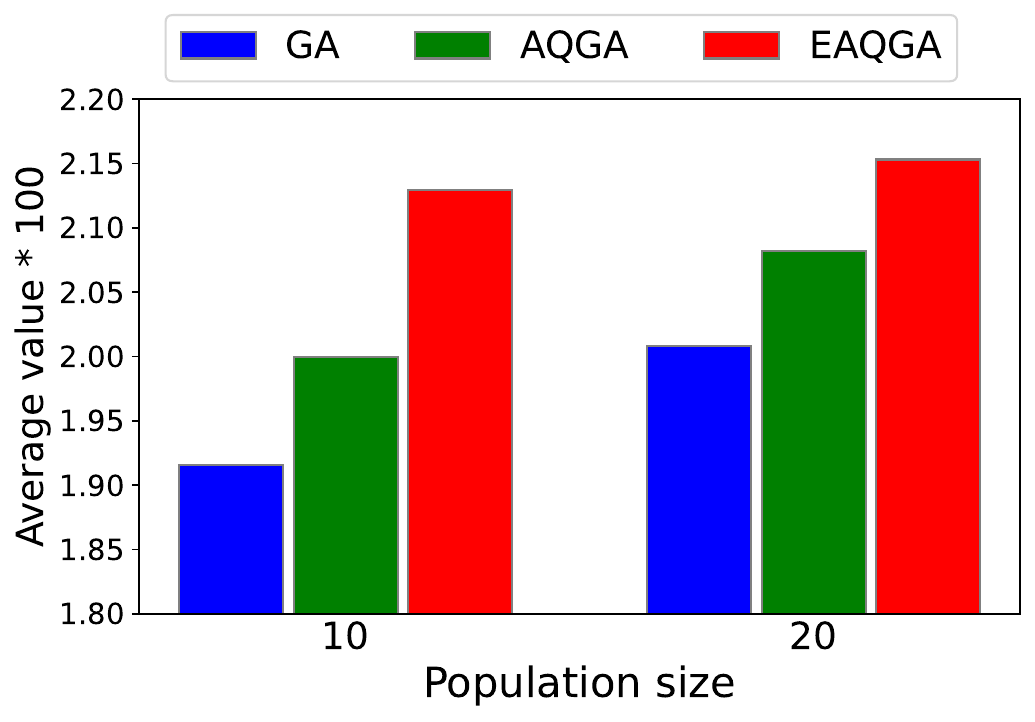}
        \label{fig:size30}
    }
    % \hspace{0.001\textwidth} % Adjust the space between the figures
    \subfigure[Size 40]{
        \includegraphics[width=0.225\textwidth]{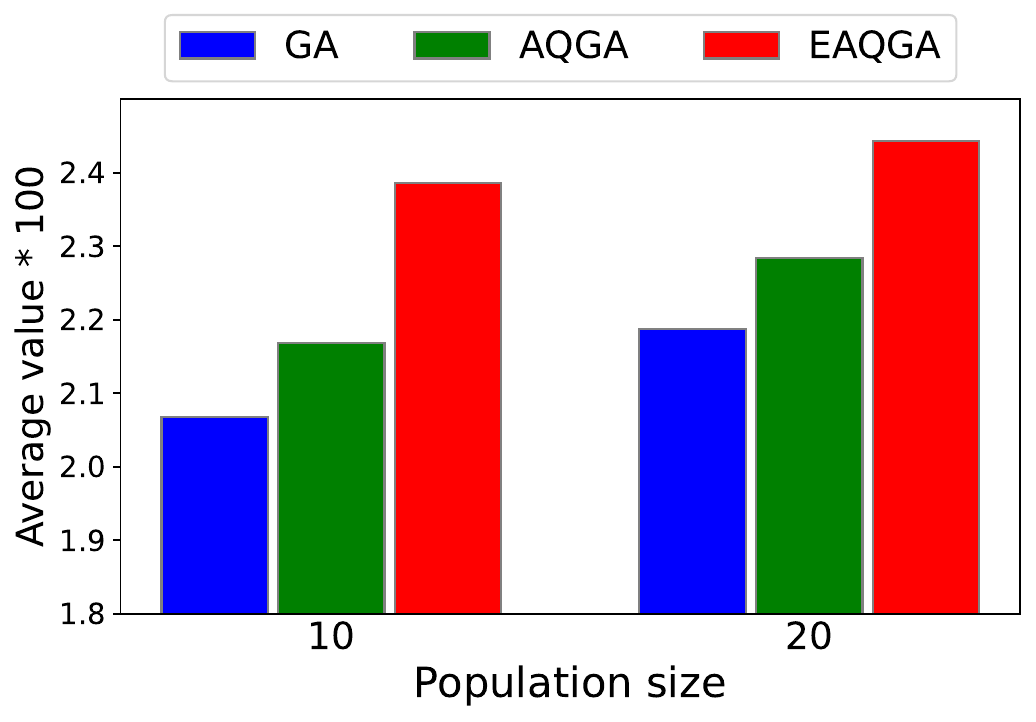}
        \label{fig:size40}
    }
    \caption{Average fitness value of three algorithms using different sizes and populations over all subsets on the simulator.}
    \label{fig:sim-avg}
\end{figure}

Figure \ref{fig:sim-avg} illustrates the average performance across all subsets, sizes, and population configurations. EAQGA consistently outperformed both GA and AQGA in every scenario. Notably, for a problem size of 40 and a population size of 10, EAQGA surpassed GA and AQGA by 15.4\% and 10\% on average, respectively. Between the latter two, AQGA performed better than GA. 
\par
Although all algorithms performed better with larger population sizes, the advantage of EAQGA was most pronounced with smaller populations (size 10), where it achieved results better than GA and AQGA with larger populations.

\subsubsection{On real quantum computer}\label{sec:real}
In this section, we present experiments conducted on \texttt{ibm\_quebec}. For this phase, we selected a significantly larger dataset subset comprising 100 stocks randomly chosen from the S\&P 500 index. Simulating 100 qubits on classical simulators would be intractable, making this a challenging yet representative problem size for industrial applications.
\par

\begin{figure}[t]
    \centering
    \includegraphics[width=0.48\textwidth]{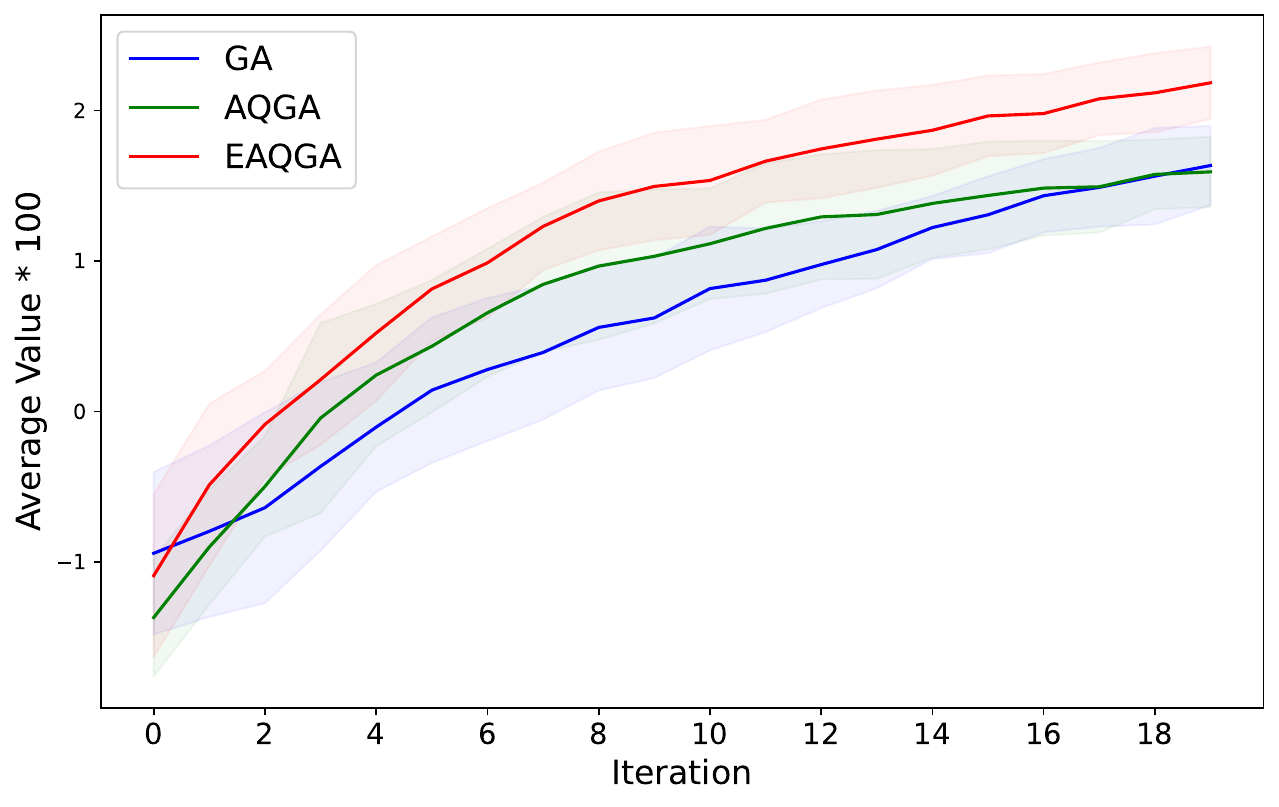}
    \caption{Average fitness over 10 runs on \texttt{ibm\_quebec}, using 100 qubits and a population size of 10.}
    \label{fig:size100}
\end{figure}

The experiments were repeated 10 times. Figure \ref{fig:size100} shows the average fitness value obtained by each algorithm at every iteration. In the first iteration, the results were entirely random, as all three algorithms started from equal superposition or random bits. Although they initially had similar average values, EAQGA outperformed GA and AQGA within a few iterations and maintained its lead through the final iteration (20\textsuperscript{th} iteration). 
\par
EAQGA produced the average fitness values that were 33.6\% better than GA and 37.2\% better than AQGA. 
\par
Given the immense search space of $2^{100}$ possibilities, EAQGA demonstrates potential for finding superior solutions compared to its classical counterparts.

\subsection{Discussion}

As demonstrated in the previous section, our proposed QGA (EAQGA) outperforms both classical genetic algorithms and the AQGA. The primary advantage of our approach lies in its ability to harness quantum phenomena such as superposition and entanglement to enhance the optimization process.

Specifically, the entanglement pattern we introduce enables simultaneous updates to pairs of variables across generations. When a pair of variables exhibits an entanglement pattern in the best solutions found so far, this relationship is preserved in the subsequent generation, although with different values. This mechanism is particularly beneficial in portfolio optimization. Such encoding allows the crossover operation to effectively capture the joint behavior of two correlated variables (qubits), maintaining their correlation across generations.

For example, if two assets exhibit a large positive covariance, similar expected returns, and a sum of returns that approximately equals the covariance divided by a risk factor ($q$), while also having low individual variances, the optimal solution may involve selecting only one of these assets. This scenario is naturally encoded by the $|\psi^n\rangle$ state in our crossover. Conversely, when two assets are highly negatively correlated, have similar negative returns, and meet similar covariance and variance conditions, it may be optimal to either include both or exclude both from the portfolio. This case is effectively represented using the $|\psi^p\rangle$ state.

In contrast to other QGAs that employ full entanglement strategies—such as those based on amplitude amplification—our algorithm utilizes a limited number of entangled pairs. This choice is motivated by current hardware constraints, as entangling gates are among the noisiest operations in today’s quantum computers \cite{ahsan2022quantum}. Moreover, our design keeps the quantum circuit depth shallow, which helps reduce noise and improve overall performance, making the algorithm more suitable for implementation on Noisy Intermediate-Scale Quantum (NISQ) devices.

It is important to note that we do not claim a quantum advantage in the formal sense, as the number of entanglements is limited and the algorithm can still be simulated efficiently on classical hardware. Nevertheless, our work introduces a meaningful enhancement to genetic algorithms inspired by quantum computing principles, offering a hybrid approach that bridges classical methods with emerging quantum capabilities.

\section{Conclusion}\label{sec:conclusion}
In this work, we introduced a new quantum genetic algorithm, EAQGA, for solving combinatorial optimization problems by generating binary solutions using quantum circuits. The core innovation lies in a novel crossover mechanism that constructs circuits for the next generation based on the best solutions from the current generation, encoding observed entanglement patterns. This crossover leverages correlated bit pairs to design circuits that increase the likelihood of reproducing correlation structures in future solutions.
\par
We evaluated this algorithm on the practical example of portfolio optimization and demonstrated its superior performance compared to GA and AQGA across the dataset. Furthermore, we experimented with portfolios of size 100 on actual quantum computers, showing that our algorithm outperforms GA and AQGA within the same number of iterations.
\par
This work provides promising insights into achieving utility-scale quantum computing with qubit size of 100, highlighting the potential advantages of quantum computing over classical approaches.
\par
Furthermore, this work will stimulate research into better understanding how quantum entanglement can be harnessed to represent various levels of correlation between assets, even in low depth circuits.

\section*{Acknowledgment}
We acknowledge access to the \texttt{ibm\_quebec} system for the experiments through $\text{PINQ}^\text{2}$. This work will be integrated into the Optimization QKit of the $\text{PINQ}^\text{2}$ innovation platform.

\bibliographystyle{ieeetr}
\balance
\bibliography{manuscript}

\begin{thebibliography}{10}

\bibitem{wolsey2014integer}
L.~A. Wolsey and G.~L. Nemhauser, {\em Integer and combinatorial optimization}.
\newblock John Wiley \& Sons, 2014.

\bibitem{hromkovivc2013algorithmics}
J.~Hromkovi{\v{c}}, {\em Algorithmics for hard problems: introduction to combinatorial optimization, randomization, approximation, and heuristics}.
\newblock Springer Science \& Business Media, 2013.

\bibitem{elbeltagi2005comparison}
E.~Elbeltagi, T.~Hegazy, and D.~Grierson, ``Comparison among five evolutionary-based optimization algorithms,'' {\em Advanced engineering informatics}, vol.~19, no.~1, pp.~43--53, 2005.

\bibitem{Bub_2010}
J.~Bub, {\em Quantum computation: Where does the speed-up come from?}, p.~231–246.
\newblock Cambridge University Press, 2010.

\bibitem{farhi2014quantum}
E.~Farhi, J.~Goldstone, and S.~Gutmann, ``A quantum approximate optimization algorithm,'' {\em arXiv preprint arXiv:1411.4028}, 2014.

\bibitem{eiben2015introduction}
A.~E. Eiben and J.~E. Smith, {\em Introduction to evolutionary computing}.
\newblock Springer, 2015.

\bibitem{telikani2021evolutionary}
A.~Telikani, A.~Tahmassebi, W.~Banzhaf, and A.~H. Gandomi, ``Evolutionary machine learning: A survey,'' {\em ACM Computing Surveys (CSUR)}, vol.~54, no.~8, pp.~1--35, 2021.

\bibitem{nielsen2010quantum}
M.~A. Nielsen and I.~L. Chuang, {\em Quantum computation and quantum information}.
\newblock Cambridge university press, 2010.

\bibitem{lahoz2016quantum}
R.~Lahoz-Beltra, ``Quantum genetic algorithms for computer scientists,'' {\em Computers}, vol.~5, no.~4, p.~24, 2016.

\bibitem{wang2013improvement}
H.~Wang, J.~Liu, J.~Zhi, and C.~Fu, ``The improvement of quantum genetic algorithm and its application on function optimization,'' {\em Mathematical problems in engineering}, vol.~2013, no.~1, p.~730749, 2013.

\bibitem{ballinas2023hybrid}
E.~Ballinas and O.~Montiel, ``Hybrid quantum genetic algorithm with adaptive rotation angle for the 0-1 knapsack problem in the ibm qiskit simulator,'' {\em Soft Computing}, vol.~27, no.~18, pp.~13321--13346, 2023.

\bibitem{xiong2018quantum}
H.~Xiong, Z.~Wu, H.~Fan, G.~Li, and G.~Jiang, ``Quantum rotation gate in quantum-inspired evolutionary algorithm: A review, analysis and comparison study,'' {\em Swarm and Evolutionary Computation}, vol.~42, pp.~43--57, 2018.

\bibitem{rubio2021quantum}
Y.~Rubio, C.~Olvera, and O.~Montiel, ``Quantum-inspired evolutionary algorithms on ibm quantum experience.,'' {\em Engineering Letters}, vol.~29, no.~4, 2021.

\bibitem{udrescu2006implementing}
M.~Udrescu, L.~Prodan, and M.~Vl{\u{a}}du{\c{t}}iu, ``Implementing quantum genetic algorithms: a solution based on grover's algorithm,'' in {\em Proceedings of the 3rd Conference on Computing Frontiers}, pp.~71--82, 2006.

\bibitem{acampora2022using}
G.~Acampora, R.~Schiattarella, and A.~Vitiello, ``Using quantum amplitude amplification in genetic algorithms,'' {\em Expert Systems with Applications}, vol.~209, p.~118203, 2022.

\bibitem{preskill2018quantum}
J.~Preskill, ``Quantum computing in the nisq era and beyond,'' {\em Quantum}, vol.~2, p.~79, 2018.

\bibitem{yanakiev2024dynamic}
N.~Yanakiev, N.~Mertig, C.~K. Long, and D.~R. Arvidsson-Shukur, ``Dynamic adaptive quantum approximate optimization algorithm for shallow, noise-resilient circuits,'' {\em Physical Review A}, vol.~109, no.~3, p.~032420, 2024.

\bibitem{gaur2023noise}
B.~Gaur, T.~Humble, and H.~Thapliyal, ``Noise-resilient and reduced depth approximate adders for nisq quantum computing,'' in {\em Proceedings of the Great Lakes Symposium on VLSI 2023}, pp.~427--431, 2023.

\bibitem{ahsan2022quantum}
M.~Ahsan, S.~A.~Z. Naqvi, and H.~Anwer, ``Quantum circuit engineering for correcting coherent noise,'' {\em Physical Review A}, vol.~105, no.~2, p.~022428, 2022.

\bibitem{vidal2003efficient}
G.~Vidal, ``Efficient classical simulation of slightly entangled quantum computations,'' {\em Physical review letters}, vol.~91, no.~14, p.~147902, 2003.

\bibitem{holland1992genetic}
J.~H. Holland, ``Genetic algorithms,'' {\em Scientific american}, vol.~267, no.~1, pp.~66--73, 1992.

\bibitem{razali2011genetic}
N.~M. Razali, J.~Geraghty, {\em et~al.}, ``Genetic algorithm performance with different selection strategies in solving tsp,'' in {\em Proceedings of the world congress on engineering}, vol.~2, pp.~1--6, International Association of Engineers Hong Kong, China, 2011.

\bibitem{han2000genetic}
K.-H. Han and J.-H. Kim, ``Genetic quantum algorithm and its application to combinatorial optimization problem,'' in {\em Proceedings of the 2000 congress on evolutionary computation. CEC00 (Cat. No. 00TH8512)}, vol.~2, pp.~1354--1360, IEEE, 2000.

\bibitem{montiel2019quantum}
O.~Montiel, Y.~Rubio, C.~Olvera, and A.~Rivera, ``Quantum-inspired acromyrmex evolutionary algorithm,'' {\em Scientific reports}, vol.~9, no.~1, p.~12181, 2019.

\bibitem{zhang2021implementation}
K.~Zhang, P.~Rao, K.~Yu, H.~Lim, and V.~Korepin, ``Implementation of efficient quantum search algorithms on nisq computers,'' {\em Quantum Information Processing}, vol.~20, pp.~1--27, 2021.

\bibitem{buonaiuto2023best}
G.~Buonaiuto, F.~Gargiulo, G.~De~Pietro, M.~Esposito, and M.~Pota, ``Best practices for portfolio optimization by quantum computing, experimented on real quantum devices,'' {\em Scientific Reports}, vol.~13, no.~1, p.~19434, 2023.

\bibitem{markowitz2000mean}
H.~M. Markowitz and G.~P. Todd, {\em Mean-variance analysis in portfolio choice and capital markets}, vol.~66.
\newblock John Wiley \& Sons, 2000.

\bibitem{gunjan2023brief}
A.~Gunjan and S.~Bhattacharyya, ``A brief review of portfolio optimization techniques,'' {\em Artificial Intelligence Review}, vol.~56, no.~5, pp.~3847--3886, 2023.

\bibitem{markowitz1952modern}
H.~Markowitz, ``Modern portfolio theory,'' {\em Journal of Finance}, vol.~7, no.~11, pp.~77--91, 1952.

\bibitem{cesarone2015linear}
F.~Cesarone, A.~Scozzari, and F.~Tardella, ``Linear vs. quadratic portfolio selection models with hard real-world constraints,'' {\em Computational Management Science}, vol.~12, pp.~345--370, 2015.

\bibitem{erwin2023meta}
K.~Erwin and A.~Engelbrecht, ``Meta-heuristics for portfolio optimization,'' {\em Soft Computing}, vol.~27, no.~24, pp.~19045--19073, 2023.

\bibitem{gunjan2024quantum}
A.~Gunjan and S.~Bhattacharyya, ``Quantum-inspired meta-heuristic approaches for a constrained portfolio optimization problem,'' {\em Evolutionary Intelligence}, pp.~1--40, 2024.

\bibitem{annealing-phase}
M.~K. Haghighi and N.~Dimopoulos, ``An enhanced hybrid approach using d-wave's cqm to solve the phase unwrapping problem,'' in {\em 2024 IEEE International Conference on Quantum Computing and Engineering (QCE)}, vol.~01, pp.~443--449, 2024.

\bibitem{QiskitPrimitives}
I.~Quantum, ``Qiskit primitives api documentation.'' \url{https://docs.quantum.ibm.com/api/qiskit/primitives}, 2024.
\newblock Accessed: 2024-10-30.

\bibitem{bayesian-optimize}
J.~Snoek, H.~Larochelle, and R.~P. Adams, ``Practical bayesian optimization of machine learning algorithms,'' in {\em Proceedings of the 26th International Conference on Neural Information Processing Systems - Volume 2}, NIPS'12, (Red Hook, NY, USA), p.~2951–2959, Curran Associates Inc., 2012.

\bibitem{sp5002024}
{S\&P Dow Jones Indices}, ``{S\&P 500 Index}.'' \url{https://www.spglobal.com/spdji/en/indices/equity/sp-500/#overview}, 2024.
\newblock Accessed: 2025-01-13.

\bibitem{qiskit_stock_data}
IBM, ``{Loading and Processing Stock-Market Time-Series Data}.'' \url {https://qiskit-community.github.io/qiskit-finance/tutorials/11_time_series.html#Loading-and-Processing-Stock-Market-Time-Series-Data}, 2024.
\newblock Accessed: 2024-10-30.

\bibitem{yahoo_finance}
Yahoo, ``{Yahoo Fianance}.'' \url {https://www.yahoo.com/author/yahoo--finance/}.

\bibitem{qiskit_mps_tutorial}
IBM, ``{Matrix Product State Simulation Method}.'' \url {https://qiskit.github.io/qiskit-aer/tutorials/7_matrix_product_state_method.html}, 2024.
\newblock Accessed: 2024-10-30.

\bibitem{schollwock2011density}
U.~Schollw{\"o}ck, ``The density-matrix renormalization group in the age of matrix product states,'' {\em Annals of physics}, vol.~326, no.~1, pp.~96--192, 2011.

\bibitem{IMB_PINQ}
IBM, ``{IBM and PINQ² unveil utility-scale quantum computer in Québec}.'' \url {https://research.ibm.com/blog/ibm-pinq2-quantum-computer-install}.

\end{thebibliography}

\end{document}